# Flexible and tunable silicon photonic circuits on plastic substrates

Yu Chen[1], Huan Li[1], and Mo Li[1*]

[1]*Department of Electrical and Computer Engineering, University of Minnesota, Minneapolis, MN 55455, USA*

**Flexible microelectronics has shown tremendous promise in a broad spectrum of applications, especially those that cannot be addressed by conventional microelectronics in rigid materials and constructions[1-3]. These unconventional yet important applications range from flexible consumer electronics to conformal sensor arrays and biomedical devices. A recent successful paradigm shift in implementing flexible electronics is to physically transfer and bond highly integrated devices made in high-quality, crystalline semiconductor materials on to plastic materials[4-8]. Here we demonstrate a flexible form of silicon photonics on plastic substrates using the transfer-and-bond fabrication method. Photonic circuits including interferometers and resonators have been transferred onto flexible plastic substrates with preserved functionalities and performance. By mechanically deforming the flexible substrates, the optical characteristics of the devices can be tuned reversibly over a remarkably large range. The demonstration of the new flexible photonic system based on the silicon-on-plastic (SOP) material platform could open the door to a plethora of novel applications, including tunable photonics, optomechanical sensors and bio-mechanical and bio-photonic probes.**

The transfer-and-bond method has been successfully implemented to fabricate flexible microelectronics and improved further with innovative mechanical designs to achieve stretchable and even foldable devices[5,8], spawning many unprecedented applications, most notably, in bio-inspired and implantable biomedical devices[9]. The new hybrid form of flexible microelectronics combines the best properties of two material worlds: the high electrical performance of crystalline inorganic semiconductor materials with the mechanical flexibility and the bio-compatibility of organic ones. Sophisticated analog and digital CMOS circuits can be transferred from silicon wafer substrates to a variety of polymeric films and retain their electrical performance and functionality in the flexible form and under mechanical deformation[7]. Beyond silicon microelectronics, the hybrid approach of flexible devices has been successfully applied to a wide range of micro-devices in diverse materials, including III-V electronics[10], microwave electronics[11,12], carbon electronics[13-16], optoelectronics[17-19], and very recently plasmonics[20,21] and meta-materials[22].

---

* Corresponding author: moli@umn.edu

Silicon photonics will enjoy the success of flexible microelectronics if they can also be transformed into a flexible form[23]. The transfer-and-bond approach is particularly viable and appealing to silicon photonics because crystalline silicon has superior optical properties, including a high refractive index and low optical loss, which are not attainable with plastic materials. There are several additional considerations that make the prospect of flexible integrated photonics uniquely promising. First, the path of light can be bent when it is guided in optical fibers or waveguides. Although glass fibers typically can only be bent to a radius of 1 cm before incurring significant loss, thanks to silicon's high refractive index (n=3.5), silicon waveguides can make a turn with a radius as small as a few microns without significant loss[24,25]. Second, unlike electronic devices, optical devices can be coupled with each other without being in physical contact — light can propagate through transparent material to couple multiple layers of optical devices. This attribute of contact-free connection could enable three-dimensional integration of photonic systems. Third, there are abundant compliant and patternable plastic materials with low refractive index and low optical absorption that are suitable for optical applications, including elastomer such as polydimethylsiloxane (PDMS), polyester such as PET (polyethylene terephthalate) and PEN (polyethylene naphthalate), and epoxies such as SU-8. Finally, the concern of silicon devices' mechanical fragility and structural stability can be addressed by the same mechanical design strategies developed for flexible electronics[26].

In this work, we demonstrate a simple yet reliable method to transfer and bond highly integrated and functional silicon photonic circuits from standard wafer substrates to flexible plastic substrates (Fig. 1a), and retain their optical performance as on the original rigid substrates. The fabrication processes are illustrated in Fig. 1b and described in detail in the Method section. Briefly, silicon photonic circuits are first patterned in the conventional way using electron beam lithography and plasma dry etching. Subsequently, the substrate is chemically etched for a precise period of time to etch the buried oxide (BOX) layer and critically undercut the silicon device layer. The undercut reduces the interfacial area between the silicon and the BOX layers so that the total bonding force between them is weakened. It is important to note that, because they are still affixed to the substrate after the etching, the devices will not move during this step. Next, a PDMS film is carefully laminated onto the substrate and peeled off at a constant speed. When the peeling speed is sufficiently high, the PDMS-silicon adhesion force can be sufficiently strong to overcome the total bonding force at the silicon-BOX interface[27], which has been reduced to the minimal by the previous etching step. In this way, the whole silicon photonic layer is lifted off from the substrate and transferred on to the flexible PDMS film. Because no adhesive material is used in the procedure, contamination to the photonic devices and consequent adverse effects on their optical performance is minimal. The strong bonding force between silicon and PDMS surfaces, however, can ensure high-yield transfer with low occurrence of dislocations and deformations. Fig. 1c and d shows optical microscope images of typical photonic circuits, including Mach-Zehnder interferometers (MZI) and micro-ring add-drop filters (ADF), after being transferred on to PDMS films. The devices consist of single-mode silicon waveguides (width, 500 nm; thickness, 220 nm) with a total length as long as 1 centimeter — an aspect ratio of $2 \times 10^4$. As shown in the images, deformations and dislocations are hardly noticeable in the transferred devices. Most notably, high magnification images in Fig. 1c and d (lower panels) reveal that the coupling gaps between the waveguides as small as 100 nm wide are precisely preserved in the fabrication process.

To characterize the optical performance of the transferred photonic devices, fiber butt-coupling method was used to couple light from a tunable laser source into the devices and to collect optical output signals to a photodetector. Figure 2a and b show typical transmission spectra of the transferred MZI and micro-ring ADF devices. The spectra measured at the two output ports of the MZI device are completely complementary to each other with a high extinction ratio. Similarly, the output spectra at the "through" port and the "drop" port of the micro-ring filter are also complementary. Those indicate that the optical coupling between the waveguides on the PDMS substrate remains efficient with a very low loss, in agreement with the observed uniform coupling gap. Fig. 2c displays broadband transmission spectrum of a critically coupled ring resonator, showing a group of resonances with the highest extinction ratio of 25 dB. From the measured quality factors Q of the ring resonators, the propagation loss in the transferred waveguide can be determined. Fig. 3d shows an under-coupled resonance at 1593.55 nm with a waveguide loaded Q of $9.9 \times 10^4$. It corresponds to an intrinsic Q of $1.5 \times 10^5$ and a propagation loss of 3.8 dB/cm. This value of propagation loss is comparable to that of the original silicon waveguides on a SOI substrate, which typically is in the range of 3-4 dB/cm if no special fabrication optimization is used[24]. It is unlikely that optical absorption in the PDMS substrate causes an increase of loss because of its transparence in the near-infrared spectral range[28,29]. We attribute any excessive loss to oxide residue and other contaminations on the waveguide surface, which can be etched away or reduced with improved transfer process. The above results demonstrate that the transfer method developed here preserves the optical performance and functionalities of the silicon photonic devices on the new plastic substrate.

Tunable photonic devices are highly desirable for applications in optical network systems that can be frequently reconfigured[30]. Conventional tuning methods either use the electro-optical effects in non-silicon materials such as lithium niobate $(LiNbO_3)$[31,32], which is difficult to integrate with silicon devices, or rely on the thermo-optical effect by electrically heating the devices[33,34]. The heating method, although integratable, needs to continuously consume electrical power to maintain the tuning. Because optical characteristics of the flexible devices apparently will change when the substrate is deformed, their functionalities can be precisely tuned by applying a controlled force, using a piezoelectric actuator for instance. Below the yield limit of the substrate material which is significantly higher for plastic materials (~50% for PDMS) than for crystalline materials (less than 1% for crystalline silicon), the device structure will respond elastically to the applied force, and reversible and reliable tuning can be achieved over a large range.

To demonstrate the tunability, the flexible photonic devices were mounted on a precision mechanical stage that can apply compression on the devices. Fig. 3 shows the results of tuning a Mach-Zehnder interferometer device when a compressive force is applied in the direction normal to the horizontal waveguides in the interferometer arms (Fig. 3a). As shown in Fig. 3b, when the substrate is compressed in steps to a strain level of 3%, the output interference fringes of the MZI shift continuously toward shorter wavelengths by 12 nm, more than one free-spectral range (FSR=10 nm). When the compression is gradually released, the fringes recover to the initial positions precisely, as marked by the vertical guidelines in Fig. 3b. Fig. 3c displays the transmission measured at two fixed wavelengths during tuning, showing sinusoidal changes with the applied strain. The transmission can be tuned between 0 and 1 when the substrate is compressed to a strain level of 1.1%. Similarly, as shown in Fig. 3d, the wavelengths of two fringe peaks shift linearly with the applied compressive strain. The transmission of the MZI is given by $T_{\text{o}} = 1/2 + \cos(\Delta\phi)/2$, where $\Delta\phi = 2\pi(n_{\text{eff}}L)/\lambda$ is the phase difference, $n_{\text{eff}}$ is the waveguide mode index and $L$ is the geometric length difference between the two interferometer arms. From the observed blue shift of the interference fringes and the sinusoidal variation of transmission in Fig. 3, it can be determined that the phase difference $\Delta\phi$ is tuned linearly with an efficiency of 163° per 1% compressive strain (or π per 1.1% compressive strain). Because the elastic modulus of silicon (130 GPa) is five orders of magnitudes larger than that of PDMS (0.3-0.7 MPa), when the applied compressive strain is above a threshold level, the silicon waveguides buckle with the PDMS substrate along the direction of applied strain (x-axis in Fig. 3a). Plentiful mechanics models have been developed to explain such a buckling effect observed in similar composite structures, mostly in the context of flexible microelectronics, and can be applied here[35-37]. The buckling amplitude $A$ is given by $A = h\sqrt{-\varepsilon_{\text{a}}/\varepsilon_{\text{c}} - 1}/(1+0.84\varepsilon_{\text{a}})$, where $h$=0.22 µm is the thickness of the silicon layer and $\varepsilon_{\text{a}}$ is applied strain (negative for compressive strain). $\varepsilon_{\text{c}} = (3\bar{E}_{\text{s}}/\bar{E}_{\text{f}})^{2/3}/4$ is the critical strain above which buckling happens. In the silicon/PDMS composite, the plain-strain modulus are $\bar{E}_{\text{f}} = 140$ GPa for silicon and $\bar{E}_{\text{s}} = 2.3$ MPa for PDMS, thus $\varepsilon_{\text{c}}$ equals 0.03% which is smaller than the minimal strain (~0.1%) that can be reliably applied in our experiment. Therefore, during the tuning, the waveguides along the direction of applied strain always buckle. The buckling amplitude $A$ at the maximal compressive strain (-3%) applied in the tuning experiment is calculated to be 2.1 µm. Since the geometric length of the waveguide increases when it buckles, the observed decrease in the phase difference $\Delta\phi$ can only be attributed to the reduction of the waveguide mode index $n_{\text{eff}}$ from the photo-elastic effect of silicon[38]. Detailed analysis in the supplementary information reveals that $n_{\text{eff}}$ of the fundamental TE mode of the waveguide along the direction of strain decreases by $\Delta n_{\text{eff}} = \eta \cdot n^3 \left[ -\rho_{12} + (\rho_{11} + \rho_{12})\upsilon \right] \varepsilon_{\text{xx}}/2$, where $n$, $\rho_{11}$ and $\rho_{12}$, $\upsilon$ are silicon's refractive index, elasto-optic coefficients and Poisson ratio, respectively. $\eta = 1.15$ is the proportional coefficient that relates the change of the waveguide mode index and the change of the material refractive index and can be determined by simulation. $\varepsilon_{\text{xx}}$ is the average normal strain in the buckled waveguide, which is tensile (positive) and can be expressed analytically in an approximate form (supplementary information) or determined numerically by simulation. The results of the theoretical model are plotted in Fig. 3c and d, showing good agreement with the experimental results with small discrepancy which can be attributed to the imprecision of the manual tuning setup used in the experiment.

The effect of mechanical tuning on the micro-ring resonators is quite different from that of Mach-Zehnder interferometers. As shown in Fig. 4b, when the sample is compressed with strain up to 9%, the resonance peaks only shift slightly by ~0.25 nm toward shorter wavelengths, about one sixteenth of the free spectral range (4 nm) of the micro-rings. In contrast, as displayed in Fig. 4c and d, both the extinction ratio and the Q factor of the ring change rapidly with applied strain. When the sample is compressed by 4%, the extinction ratio first increases from 3 dB to a maximal value of 22 dB, indicating the resonator reaches the critical coupling condition. At the same time, the loaded Q factor increases from $5 \times 10^3$ to $1.5 \times 10^4$, suggesting that the intrinsic Q factor of this micro-ring device is ~$3.0 \times 10^4$. Further compression reduces the extinction ratio until the resonances disappears while the Q factor continues to rise to $2.5 \times 10^4$, approaching the intrinsic value. The tuning behavior can be explained by the increase of the gap between the waveguide and the micro-ring when the substrate is compressed. Compressing the substrate causes buckling of the film and consequently lateral and vertical offset between the waveguides (Fig. 4a). This increase in the coupling gap reduces the coupling coefficient $\kappa$ and causes the waveguide-ring system to be tuned gradually from the initial over-coupled condition to critically-coupled and further to under-coupled conditions. Analysis using standard theory of optical resonators[39] (lines in Fig. 4c and d) indicates that the effective coupling gap is tuned from the initial value of 80 nm to about 112 nm to reach critical coupling when 3.7% compressive strain is applied. On the other hand, the resonator's weak spectral sensitivity to mechanical tuning can be understood possibly from the relaxed strain in a ring structure and the symmetry of the photo-elastic effect in a closed loop under uniaxial strain, although a more complete mechanics model is still needed to analyze the detailed distribution of strain field in the ring.

The demonstrated ability to tune the optical properties of the flexible silicon photonic devices over such a large range will find important applications in adaptive and reconfigurable optical systems. In addition, the devices' sensitive response to substrate deformation implies they can be applied as optomechanical sensors to measure mechanical load and displacement with high sensitivity. Their flexible format allows them to be "taped" conformally on curved surfaces such as on animal and human skin. The flexible devices demonstrated here are mechanical robust and the tunability is reversible and repeatable. We have conducted tests in which the devices were tuned repeatedly for more than fifty cycles. The results show that the optical properties of devices can be recovered to within 2% range of the original value. The small unrepeatability can be attributed to the imprecision of the manual tuning setup and the effects of friction. The devices do not fail until they are deformed to a very large extent of more than 20% deformation. The failure mechanisms include cracking, slipping and delamination of the silicon layer from the substrate. To further improve the devices' mechanical robustness, mechanical design strategies such as using additional adhesive layers or placing the devices at the strain neutral plane of a multilayer film can be employed[26].

The demonstration of transferring flexible silicon photonic devices onto plastic substrates, their preserved optical functionalities, mechanical resilience and tunability, is a significant first step toward a fully integrated flexible photonic system. The devices on PDMS substrate can be subsequently transferred onto a variety of plastic materials. By advancing the method demonstrated here and those developed in flexible electronics research and solving the challenge of precise alignment, it will be possible to assemble multiple layers of flexible silicon photonic devices with active optical devices made of non-silicon material (such as germanium and III-V semiconductors) in three dimensions. A complete photonic system thus can be realized, leading toward a wide range of future applications that require mechanical flexibility and biocompatibility, including implantable biophotonic sensors and optogenetic probes.

**Method**

The silicon photonic circuits were first patterned on a standard silicon-on-insulator wafer (SOITEC, Unibound, 220 nm top silicon layer, 3 μm buried oxide layer) using one-step electron beam lithography (Vistec, EBPG 5000+) and plasma dry etching (Trion II ICP-RIE) with chlorine based chemistry. Subsequently, the substrate was etched in 10:1 buffered oxide etch (BOE) solution, for a precise period of time and at a precisely controlled temperate, to etch the buried oxide (BOX) layer and undercut the silicon device layer. The undercut reduces the interfacial area between the silicon and the BOX layers so that the total bonding force between them is reduced to the minimal without separating them. After etching, the silicon device layer is still affixed to the substrate so that they will not move. In the next step, a PDMS film was laminated onto the substrate. The PDMS film was made from Sylgard 184 (Dow Corning, Inc.) mixture with 10:1 ratio and baked at 90 degree Celsius for 1 hour. The film was first thoroughly cleaned with isopropyl alcohol (IPA) and dried in nitrogen gas flow. UV-induced ozone (Jelight UVO-Cleaner) was then used for two minutes to treat the surfaces of the substrate and the PDMS film, allowing strong, covalent bonding to form between them when they are in physical contact. Uniform mechanical pressure and adequate desiccation were applied to ensure conformal contact between the two surfaces and to release water moisture trapped at the interface. The ends of the waveguides were aligned to the edge of the PDMS film to allow fiber butt-coupling in the measurement. Finally, the PDMS film was manually peeled off from the substrate at a constant speed. With sufficiently high peeling speed, the PDMS-silicon adhesion force is adequate to overcome the total bonding force at the silicon-BOX interface, which has been reduced to the minimal by the previous undercut etching step. In this way, the whole silicon photonic layer was lifted off from the substrate and transferred to the flexible PDMS film. The device was then mounted on a fiber alignment stage. Two tapered fibers with 2 μm focused spot size were aligned to the ends of the transferred waveguide. Typical fiber to waveguide coupling efficiency is 10%. Mechanical tuning was realized by compressing the device using another manually controlled mechanical stage with a precision of 10 μm. The imprecision of this manual stage, as well as the misalignment of the fibers to the waveguides during the tuning, leads to the errors and some unrepeatability in the experiment.

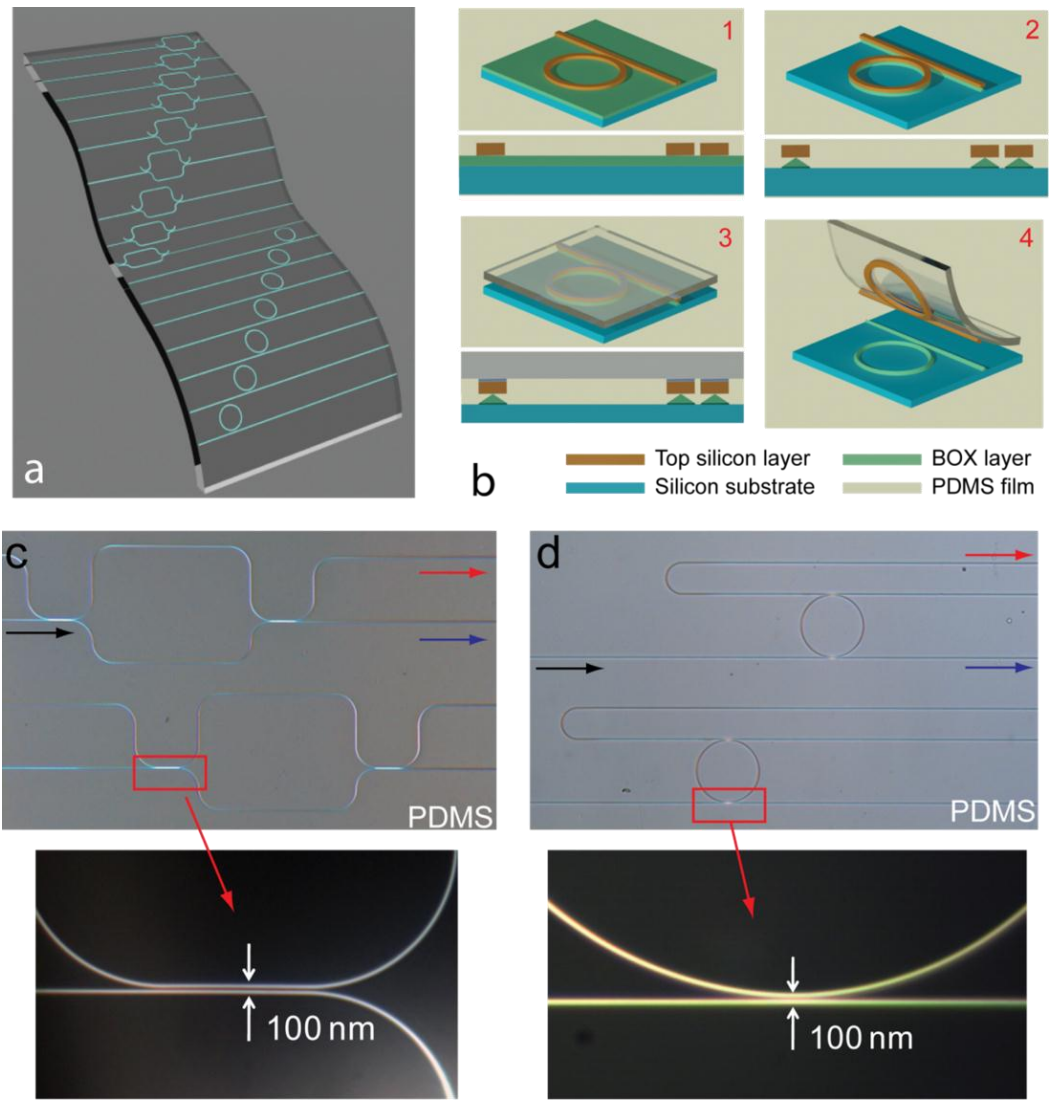

**Figure 1 Transfer and bond fabrication of flexible silicon photonic circuits. a**, Illustration of silicon photonic circuits on a flexible plastic substrate. **b**, Fabrication procedures to transfer and bond silicon photonic circuits from a wafer substrate to a flexible PDMS substrate. **c** and **d**, Optical microscope images of Mach-Zehnder interferometers (MZI) (c) and micro-ring add-drop filters (ADF) (d) after being transferred onto PDMS substrate. The high magnification images (lower panels) show that the coupling gaps of 100 nm between the waveguides are precisely preserved during the transfer.

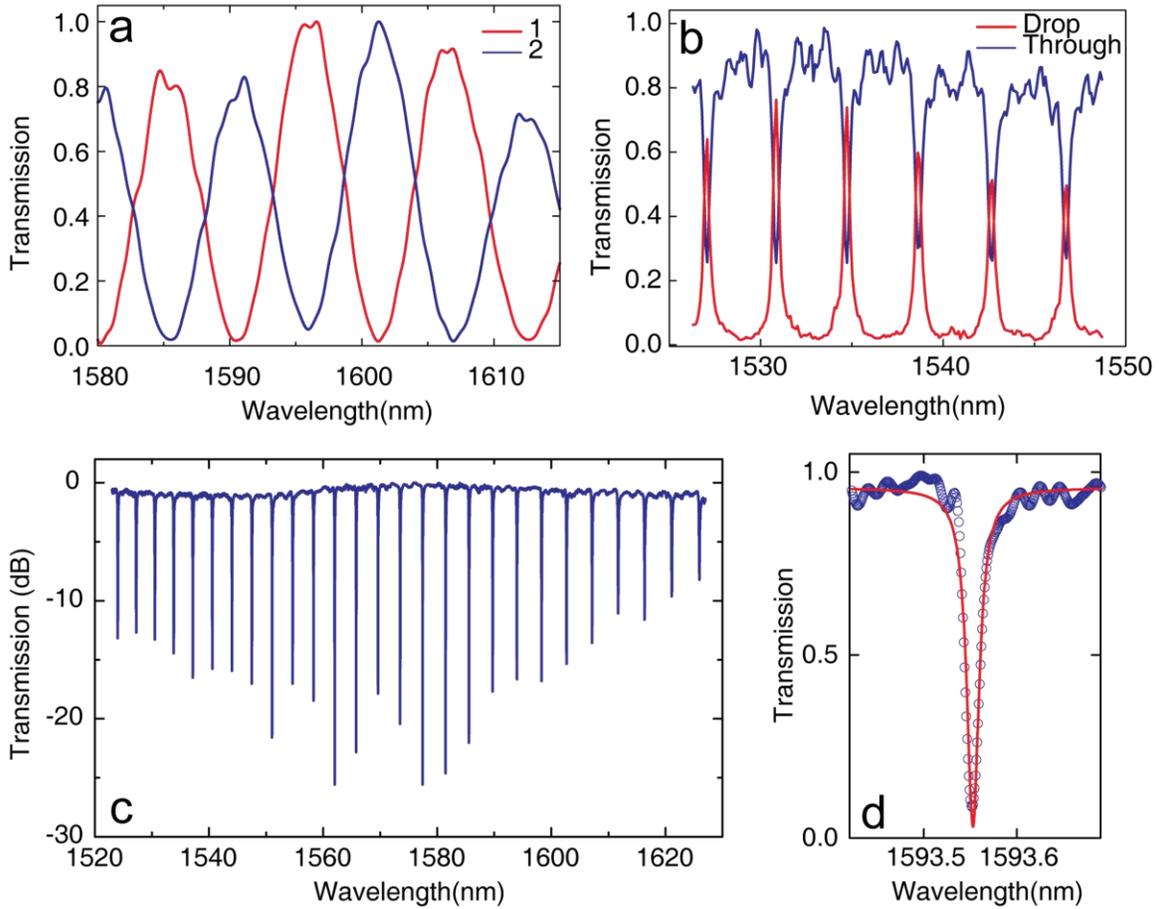

**Figure 2 Preserved optical functionalities of silicon photonic circuits on PDMS substrate. a**, The transmission spectra of the MZI circuit measured at its two output ports, showing high extinction ratio and complementary interference fringes. **b**, Transmission spectra of the micro-ring ADF measured at the "through" and "drop" ports, also showing complementary resonance peaks. **c**, Broadband transmission spectrum of a ring resonator critically coupled to a waveguide on the PDMS substrate, showing a high extinction ratio up to 25 dB. **d**, Measured high-Q resonance (blue symbols) and Lorentzian fitting (red line) of a transferred ring resonator with a loaded quality factor of $9.9 \times 10^4$ and an intrinsic quality factor of $1.5 \times 10^5$. The corresponding value of propagation loss in the waveguide is 3.8 dB/cm.

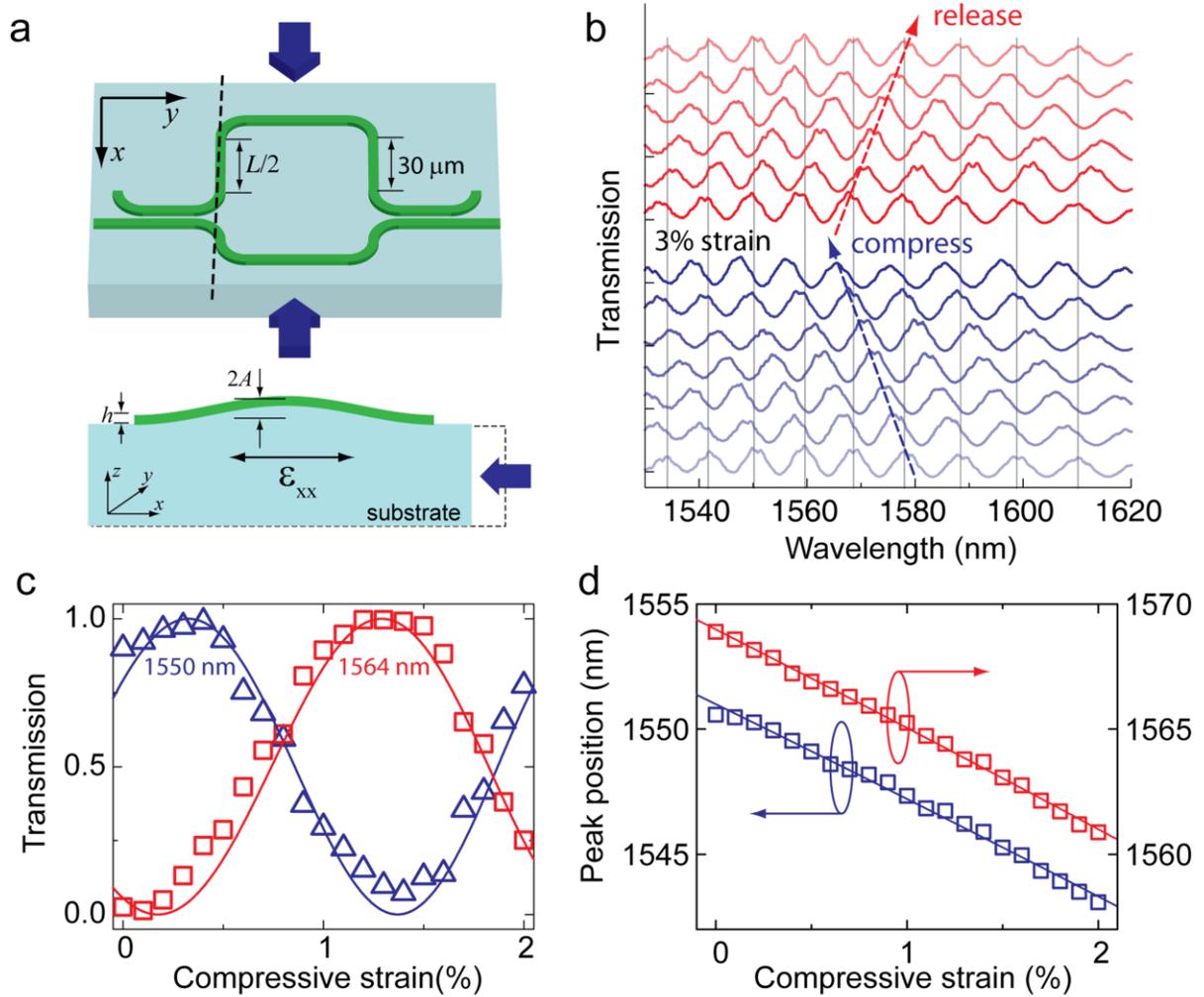

**Figure 3 Mechanical tuning of a Mach-Zehnder interferometer on an elastic substrate.**
**a**, When the substrate is compressed with strain beyond a critical value, the silicon waveguide buckles with the PDMS substrate. **b**, Under increasing compression, the interference fringes in the output of MZI continuously shift toward shorter wavelengths. When the compression is relaxed, the fringes recover to their initial spectral positions. **c**, At given wavelengths (1550nm, blue; 1564nm, red), the measured transmission (symbols) varies sinusoidally with the increasing compressive strain, in good agreement with the results of the theoretical model (lines). **d**, The peak wavelengths (symbols) of the interference fringes shift linearly with the increasing compressive strain toward shorter wavelengths as expected from the theory (lines).

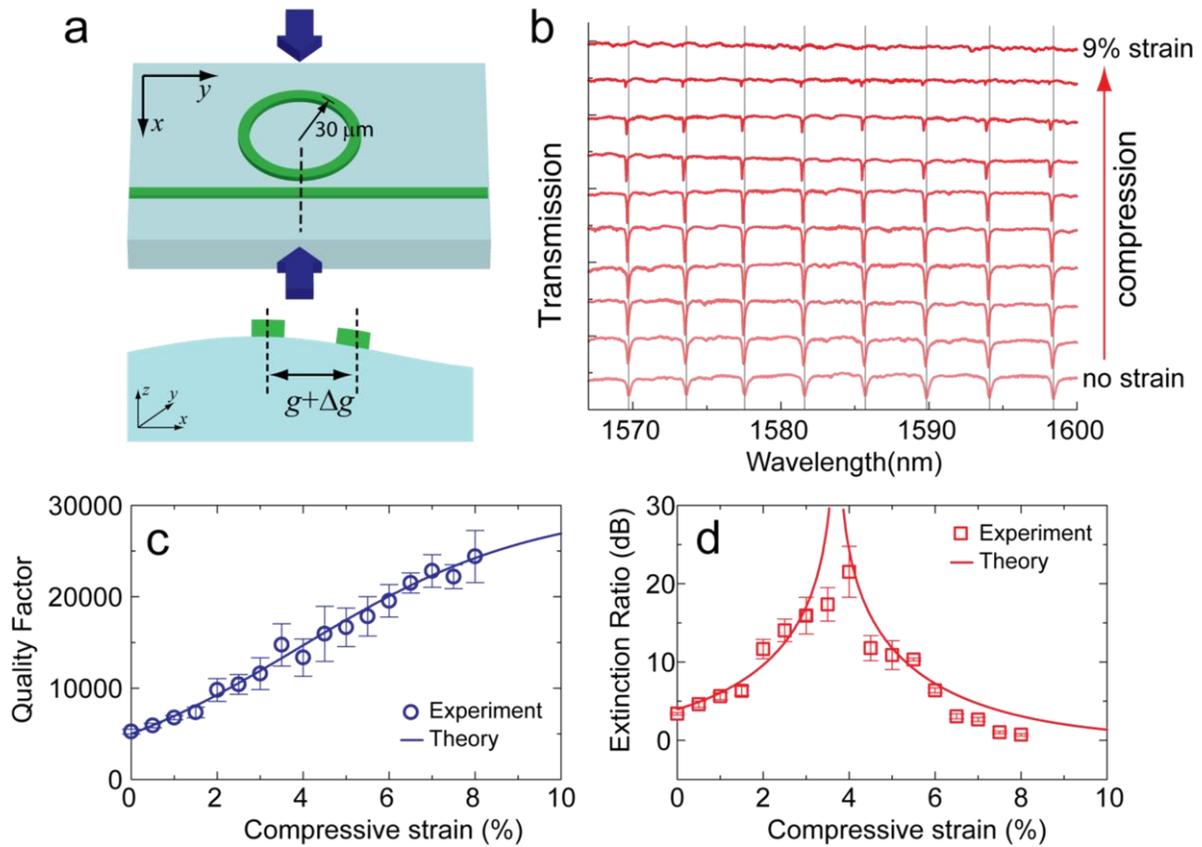

**Figure 4 Mechanical tuning of a micro-ring resonator device on an elastic substrate. a**, Compressing the substrate induces an increase in the coupling gap between the micro-ring and the coupling waveguide. **b**, Under increasing compression, the wavelengths of the resonances only shift slightly whereas the resonance extinction ratios and quality factors changes dramatically. **c** and **d**, Quality factor (c) and extinction ratio (d) versus the applied compressive strain. The quality factor increases five folds over a range of 8% strain. The extinction ratio can obtain a maximal value of 22 dB when the critical coupling condition is reached at a strain level of 3.7%. The results agree with a theoretical model (lines) assuming the coupling gap increases linearly with the applied compressive strain.